\documentstyle[11pt]{article}
\def\etal{{\it et al} \ }
\def\kms{km s$^{-1}$}

\def\gsim{ \lower .75ex \hbox{$\sim$}
\llap{\raise .27ex \hbox{$>$}} }
\def\lsim{ \lower .75ex \hbox{$\sim$}
\llap{\raise .27ex \hbox{$<$}} }
\def\pp{\noindent\parshape 2 0truecm
16.0truecm 1.0truecm 15truecm}

\def\spose#1{\hbox to 0pt{#1\hss}} \def\simlt{\mathrel{\spose{\lower
3pt\hbox{$\mathchar"218$}} \raise 2.0pt\hbox{$\mathchar"13C$}}}
\def\simgt{\mathrel{\spose{\lower 3pt\hbox{$\mathchar"218$}} ' \raise
2.0pt\hbox{$\mathchar"13E$}}} \def\oneskip{\vskip\the\baselineskip}

\begin{document}
\baselineskip 12pt
\centerline{\LARGE Galaxy Harassment and the Evolution of Clusters of Galaxies}
\oneskip
\oneskip
\centerline{\bf
Ben Moore$^1$, Neal Katz$^1$, George Lake\footnote{Department of Astronomy,
University of Washington, Seattle, WA 98195-1580, USA}, }
\vskip 0.1truecm
\centerline{\bf
Alan Dressler\footnote{Observatories of the Carnegie Institution of Washington,
Pasadena, CA 91101-1292, USA}
and
Augustus Oemler, Jr\footnote{Yale University Astronomy Department, New Haven,
Ct 06511, USA}}

\oneskip
\oneskip
\oneskip

{\bf Disturbed spiral galaxies with high rates of star formation
pervaded clusters of galaxies just a few billion years
ago, but nearby clusters
exclude spirals in favor of ellipticals.  ``Galaxy harassment"
(frequent high speed galaxy encounters) drives the morphological
transformation of galaxies in clusters, provides fuel for quasars in
subluminous hosts and leaves detectable debris arcs.  Simulated images
of harassed galaxies are strikingly similar to the distorted spirals
in clusters at $z \sim 0.4$ observed by the Hubble Space Telescope.}

\bigskip

Clusters of galaxies are unique cosmological laboratories.  There are
several hundred galaxies moving at relative velocities up to several
thousand \kms\ in regions no larger than the distance between the Milky
Way and its nearest neighbor, the Andromeda galaxy (M31).  Clusters of
galaxies have been observed at redshifts up to 2$^{\bf\rm 1}$.  By
understanding their evolution over cosmic times, we probe the geometry
of the Universe and the development of its largest structures.

Nearby rich galaxy clusters are dominated by elliptical ``E" and
lenticular ``S0" galaxies$^{\bf\rm 2}$, mostly low luminosity dwarfs.
Twenty years ago Butcher and Oemler$^{\bf\rm 3,4}$ discovered that
clusters at $z \gsim 0.4$ have a substantial population of ``blue
galaxies" seen only as fuzzy blobs in their ground based images.
Recent Hubble Space Telescope (HST) images revealed that the ``fuzzy
blue blobs" are low luminosity, often disturbed, spiral galaxies
``Sp"$^{\bf\rm 5-8}$.  The HST imaging teams stress that the disturbed
blue galaxies are ubiquitous, but few have
other galaxies nearby$^{\bf\rm 5}$ and
there were multiple bursts of star formation
spanning up to 2 Gyr$^{\bf\rm 8}$.
The dramatic transformation of clusters (shown in Figure 1)
occured during a ``look-back time" of just 4-5 billion years,
only a few cluster
orbital times.  In contrast, the morphological fraction in the field
shows far less evolution$^{\bf\rm 9}$.

\oneskip
{\bf Figure 1: \em The dramatic evolution of clusters from $z \sim
0.4$ to today is shown in these images of their inner $300h^{-1}$ kpc,
where $h$ is the Hubble constant in units of 100 \kms\ Mpc$^{-1}$.
The left panel is the nearby Coma cluster at $z=0.023$ (courtesy
A. Fruchter, B. Moore and C. Steidel).  Nearly every object
surrounding the two central dominant elliptical galaxies is an E or
S0.  The right panel is an HST image of CL0939 at $z=0.41$$^{\bf\rm
6}$.  This cluster is dominated by spiral galaxies; many appear
disturbed yet no other galaxies are nearby.  While others appear
normal, they have enhanced rates of star formation betrayed by strong
{\em H}$_\beta$ and {\em O}$II$ emission lines.  The bright elliptical
population is already in place at this redshift. The difference in
galaxy populations is most dramatic at luminosities more than two
magnitudes fainter than $L_*$, the characteristic break in the
luminosity function$^{\bf\rm 10}$. Fainter than this luminosity, 90\%
of galaxies in clusters at $z \sim 0.4$ are bulgeless ``Sd'' disk
systems, whereas 90\% of galaxies within nearby clusters are dwarf
ellipticals ``dE'' or dE/S0 $^{\bf\rm 11,12}$. The resolution of both
images is $\sim 0.5h^{-1}$ kpc. }
\oneskip

Given a mechanism for distorting galaxies and promoting star formation
that operates when a spiral first enters a cluster, hierarchical
clustering models will naturally enhance the number of
``Butcher-Oemler clusters" at $z \sim 0.4$$^{\bf\rm 13}$. Proposed
mechanisms include: mergers$^{\bf\rm 14,15}$, compression of gas in
the high pressure cluster environment$^{\bf\rm 16,17}$ and tidal
compression by the cluster$^{\bf\rm 18,19}$.  Each of these scenarios
can produce star-bursts, but none address morphological evolution or
identify the remnants of these distorted blue galaxies in present day
clusters. By analyzing their HST images, Oemler \etal$^{\bf\rm 20}$
conclude that merging is implausible as the blue galaxy fraction is
large and the merging probability is low.  They observe disturbed
spirals throughout the cluster, whereas both ram pressure stripping
and global tides will only operate efficiently near the cluster's
center.

What mechanism drives star-bursts and rapid morphological evolution
{\it throughout} a cluster of galaxies?  Although direct mergers are
extremely rare, every galaxy experiences a high speed close encounter
with a bright galaxy once per Gyr.  Here, ``close" means within 50
kiloparsecs (several optical radii) and ``bright galaxy" is one as
luminous as $L_*$, the characteristic break in the galaxy luminosity
function$^{\bf\rm 10}$.  (The Milky Way has a luminosity of
approximately $L_*$.)  We use high resolution numerical simulations of
galaxies in clusters to examine the resulting damage of these close
encounters.  To distinguish this from other collisional effects such
as galaxy mergers and galaxy cannibalism, we refer to these frequent
encounters as ``galaxy harassment''.

\section{The cluster environment and galaxy harassment}

We simulate  the evolution of a small
bulgeless spiral galaxy orbiting within a dense cluster modeled on Coma
(Figure 1).  Figure 2 shows an edge on view of the model galaxy along
with an inventory of its stars, gas and dark matter.  Some of our
simulations use smoothed particle hydrodynamics$^{\bf\rm 21}$,
to evolve the gas component of the disk at resolutions of 100 - 500 pc.

We simulated galaxies on circular and elliptical orbits in smooth
cluster potentials before examining the effects of harassment.  The
disk galaxy shows little evolution over 5 Gyrs when placed on a 450
kpc circular orbit in a smooth cluster potential.  The disk becomes
bar unstable after the first pericentric passage$^{\bf\rm 18}$ when
placed on an eccentric orbit with apocenter and pericenter at 600 kpc
and 300 kpc respectively.  Thereafter, each time the galaxy passes
through pericenter, the halo loses a small fraction of its mass but
stars and gas remain bound.  The evolution is far more dramatic after
including the other cluster galaxies.

\oneskip
{\bf Figure 2: \em An edge on view of a model Sd galaxy.  Blue, red
and yellow particles make up the dark matter, gas and stars
respectively. The exponential disk has a scale length of 2.5 kpc and
scale height of 200 pc. The disk is constructed with a Toomre
``stability'' parameter $Q=1.5$ and is run in isolation for 2 Gyr (as
shown) before being set into orbit in the cluster.  The gaseous disk
is initially on cold circular orbits.  The dark halo is a spherical
isothermal with a core radius of one kpc that is tidally truncated at
the pericenter of the galaxy's orbit.  Within 20 kpc the ratio of dark
matter to stars to gas is 20:5:1 and the total mass is $\sim
10^{11}M_\odot$.  The graph displays the contribution of each
component to the rotational velocity of the disk.  Using the
relationship found by Tully \& Fisher$^{\bf\rm 22}$, a circular
velocity of $\sim 160$ \kms \ corresponds to a galaxy with a
luminosity about $L_*/5$. We performed similar simulations using model
galaxies with a peak circular velocity of $\sim 110$ \kms.  The
galaxies are placed in a model of the Coma cluster that has a total
mass within its virialised radius of $7\times10^{14}h^{-1}M_\odot$.
Galaxies within the cluster are  drawn from a
Schechter luminosity function normalised to a typical cluster
mass-to-light ratio of $250 h$.  This results in a cluster with 950
galaxies brighter than the Magellanic clouds, but only 31 brighter
than $L_*$.  Internal velocity dispersions are assigned using the
observed correlation with luminosity$^{\bf\rm 23}$.  The assigned
masses assume that the dark halos are tidally truncated at their
pericentric distance reduced by a time averaged loss of 25\% owing to
harassment$^{\bf\rm 24}$.  The fraction of the cluster's density attached to
galaxies varies from zero at its center to nearly unity at its virial
radius of 1.5 $h^{-1}$ Mpc.  At 300 $h^{-1}$ kpc,
roughly 15\% of the mass is
bound to galaxies.  The rest of the cluster mass is in a smoothly
distributed background represented by a fixed analytic potential.
Further details can be found in Moore \etal$^{\bf\rm 24}$.}
\oneskip

When we include harassment, the havoc wreaked is determined by the
masses of bright galaxies in clusters.  Galaxies in the field have
massive dark halos, but there has been speculation that these were
stripped from individual galaxies within clusters$^{\bf\rm 25}$.  All
galaxies are tidally limited by the potential field of the cluster.
Over the 5 Gyr evolution of the cluster, bright galaxies retain more
than half of the mass within their tidal radius (measured at the
pericenter of their orbit), the rest being liberated by fast
encounters with other bright galaxies$^{\bf\rm 24}$.

We have been conservative in our simulations to insure that our
results are robust.  At a fixed mean orbital radius or ``guiding
center", the effects of harassment becomes stronger as orbits become
more elongated.  In our cluster model, the mean ratio of a perturbing
galaxy's apocenter to its pericenter is greater than 10-to-1.
A galaxy in a rich cluster with a guiding center of 450 kpc will have a
typical pericenter ($r_{peri}$) slightly greater than 150 kpc.  The
masses we assign galaxies are $\sim 2.8\times10^{11} (r_{peri}/150kpc)
(L/L_*)^{3/4} M_\odot$.  As a result, their mass-to-light ratios are
$M/L = 44 h^2  (r_{peri}/100kpc)(L/L_*)^{-1/4}$.
The luminous parts of elliptical galaxies are observed to have
mass-to-light ratios of
$12  h M_\odot/L_\odot$$^{\bf\rm 26}$, so the perturbing galaxies have modest
amounts of dark matter.
However, we follow the evolution of individual
harassed galaxies that have apo/peri ratios of 2 ({\it i.e.} apocenter
at 600 kpc, pericenter at 300 kpc).  As a result, this galaxy avoids
extremes of the cluster distribution and starts with a large dark halo
mass since its pericentric distance is not far from its guiding
center.  Both effects serve to underestimate the effects of
harassment.

\section{The changing morphologies of harassed galaxies}

In Figure 3, we follow the evolution of a harassed galaxy through its
slightly eccentric orbit.  At each phase, we compare with images of
galaxies in clusters.
Our initial conditions look like a normal spiral galaxy found in the
field (Figure 3a).  Typically, the first encounters create ``disturbed
barred spirals" with sharp and dramatic features drawn out from the
dynamically cold disk (Figure 3b).  Tails of material can be pulled
out and distorted by the tidal field of the cluster (Figure 3c).  The
gas distribution often forms ring structures that tumble within the
stellar bar (Figure 3d).

\oneskip

{\bf Figure 3: \em Comparisons of synthetic images from our
simulations with observations of harassed galaxies.  In the simulated
images, the stellar distribution is smoothed and filtered to model the
resolution of the observations.  The limiting brightness contours are
about 27 magnitudes per square arcsecond.  (a) The left panel shows
the initial model galaxy viewed face and edge on.  The size of each
image is 40 kpc across.  (b) The upper image shows a disturbed spiral
galaxy taken from the Butcher-Oemler cluster CL1447. The lower image
shows our model galaxy 150 million years after suffering a single
strong encounter that has pulled the two tails of material from the
disk. At this time, the perturbing galaxy has already moved over 200
kpc away.  (c) The upper image is NGC 4438, a spiral galaxy near the
center of the Virgo cluster (from Sandage and Bedke$^{\bf\rm 27}$).
The close companion in this image is probably not responsible for the
disturbance$^{\bf\rm 28}$. The lower image shows a snapshot of our model galaxy
after a Gyr in the cluster.  The tidal tails of material pulled from
the galaxy have been subsequently tidally distorted.  (d) The upper
image is a spiral galaxy with a prominent ring in the distant rich
cluster CL0939.  The lower image shows the ring structure often
observed in harassed galaxies.  (e) Our model galaxy after 3 Gyrs of
evolution.  The stellar distribution shown here should be compared
with the initial model shown in Figure 3a and with the dwarf
elliptical galaxies shown in Figure 3f.  (f) CCD images of dwarf
elliptical galaxies in the Coma cluster.  The pixel scale is roughly
the same physical size as shown in Figures 3(a-e).  }

\oneskip

The evolution is driven by several close encounters that would drive
the multiple starbursts inferred from HST data$^{\bf\rm 8}$.
Another observational puzzle has been the ubiquity of disturbed galaxies
with no sign of current interaction$^{\bf\rm 6}$.  None of the images
of our model galaxy has another cluster galaxy within 50 kpc.  Over
the course of 3 Gyr, the closest approach of another galaxy is more
than 30 kpc away.  Since the relative velocity of strong encounters is
$\sim 1,500$ \kms, and the velocity impulse internal to the galaxy is
only $\lsim$50 \kms, the perturbing galaxy moves $\sim 100$ kpc by the
time the disk's response is visible.

After several strong encounters, the loss of angular momentum to their
own dark halos and the perturbing galaxies, combined with impulsive
heating, leads to a prolate figure supported equally by random motions
and rotation.  The gas sinks to the very center of the galaxy and the
stellar distribution is heated to the extent that it closely resembles
a dwarf elliptical, although some retain very thick stellar disks and
would be classed as dwarf lenticulars.  At this stage in the evolution
encounters cease to create sharp distortions and fail to remove any
more material from the compact remnant.

The final stellar systems have a large degree of rotational support,
surface density profiles and shapes that are in good agreement with
observations.  Figure 3e shows the stellar configuration after 3 Gyrs
and can be compared with real images of dwarf elliptical galaxies
taken from the Coma cluster (Figure 3f).  Note that the final
photometric axes of the model galaxy are tilted with respect to the
initial plane of rotation of the disk.

Using our simulations, we can identify the present-day remnants of
the disturbed spirals seen at $z \sim 0.4$.  Below $L_*$, two distinct
classes of elliptical galaxies are observed.  Low luminosity Es with
high central surface brightness are a rare extension to the sequence
of bright ellipticals; the archetype is M32. The most numerous
galaxies in clusters are in a second class of dwarf ellipticals, also
known as dwarf spheroidals (dE/dSph).  Their exponential surface
brightness profiles resemble those of spirals, as does the correlation
of their low central surface brightnesses with total luminosity.  They
are faint, at least 3 magnitudes below $L_*$ and as many as 14
magnitudes if one extrapolates to the faintest known galaxies in the
Local Group, Draco and Ursa Minor$^{\bf\rm 29,30}$.  Harassment
transforms spirals into this latter class of galaxies.

The observed stellar populations of dE galaxies implies recent star
formation activity that can easily be understood in our model as a
result of recent encounters with cluster galaxies.  Harassed Sd spiral
galaxies undergo a remarkable transformation from one morphological
class to another without any merging taking place.  Their dynamical
states can account for all of the dissimilarities between dwarf
elliptical and normal elliptical galaxies.  Harassment provides the
link between the dominant populations of galaxies in clusters at $z
\sim 0.4 $ and the present-day.

\section{Other effects of harassment:  tidal debris and quasar fuel}

Stellar and gaseous material torn from the disk during violent
encounters creates debris tails that lead and follow the galaxy's
orbit through the cluster.  At pericenter, the tidal debris creates a
giant arc that could be mistaken for a gravitational arc when viewed
face-on (Figure 4).  However, the transverse size is slightly greater
than a lensed galaxy and its redshift will match the cluster.  The
harassment of the debris tails should create tidal shocks that promote
the formation of dwarf galaxies, as seen in observations and
simulations of tails in merging galaxies$^{\bf\rm 31,32}$.  Low
surface brightness features punctuated with dwarf galaxy formation
should be detectable in most HST images of $z \sim 0.4$ clusters.

\oneskip
{\bf Figure 4: \em The smoothed surface brightness of the stellar
tidal debris after 4 Gyrs of evolution. The image is 2 Mpc across and
the intensity of the colour shows the logarithm of the smoothed
stellar surface density plotted wherever the surface brightness
$m_b<30$ magnitudes per square arcsecond. The white dots are
individual particles from the galaxy's dark halo.  The two long tails
were stripped by strong encounters with other cluster members coupled
with the mean tidal field of the cluster.  At various positions along
the orbital path the stripped stars arrange themselves into long arcs
following the galaxy's orbit.  HST images should show the brightest
features (where the colour is brighter than orange).}
\oneskip

Future observations of intra-cluster light coupled with simulations
covering the large parameter space of orbits and luminosities will
provide interesting constraints on our evolutionary scenario.  Current
observations are conflicting, anywhere from 0\% to 30\% of the total
cluster luminosity could reside in a diffuse component$^{\bf\rm
33,34}$.  In our cluster model, 20\% of the light starts off in
galaxies fainter than $L_*/5$.  After 5 Gyrs, $\sim$20\% of the stars
are lost to the intra-cluster medium from our small Sd galaxies.  The
bulk of the evolution is driven by the few ($\lsim$ 5) strong
encounters with galaxies brighter than $\sim L_*$.  As a result, the
evolution is chaotic: whereas one fragile disk galaxy can avoid strong
encounters for a few Gyrs, another may be completely destroyed.  The
total quantity of stripped material can vary from a few percent, to
rarer cases where the entire galaxy is disrupted after several strong
encounters.

Recent HST images of quasars suggest that many quasar hosts are {\it
not} galaxies as luminous as $L_*$ $^{\bf\rm 36}$.  This is surprising
as simple energy considerations imply that quasars need $10^8M_\odot -
10^9M_\odot$ to fuel the black hole engine.  To be conservative, one
would prefer models where no more than 10\% of a galaxy's gas must be
channeled to the center.  This argues for hosts at least as large as
our own Milky Way, which has a few times $10^9M_\odot$ of gas.  From 8
HST images of low redshifts quasars, Bahcall \etal$^{\bf\rm 36}$ found that
5 of the host galaxies must be 0.5 to 1.5 magnitudes fainter than
$L_*$.  Spiral galaxies of this luminosity have gas masses of order
$10^8M_\odot$, a quasar needs nearly all of it for fuel!

Galaxy harassment can give the quasar the fuel it needs.  While 10\%
of the gas is tidally stripped from the galaxy, the remaining 90\% of
the gas ($\gsim 10^8M_\odot$) sinks to the inner few hundred parsecs---the
resolution of our simulation---by rapidly losing angular momentum to
the perturbing galaxies, dark halo and stellar bar.

Clearly, many quasars have luminous hosts$^{\bf\rm 36}$.  Also,
quasars are known to avoid rich clusters$^{\bf\rm 37}$, so how could
they be harassed?  Oddly, three of the five quasars with subluminous
hosts found by Bahcall \etal $^{\bf\rm 35}$ lie in clusters.  A fourth
is probably in a cluster and the environment of the fifth has not been
studied.  Harassed galaxies provide ideal hosts for quasars at
intermediate redshifts known to lie in subluminous galaxies$^{\bf\rm
38}$.

\section{Concluding Remarks}

Galaxies are metamorphised by their mutual interactions.  ``Merging"
of spirals in groups creates bright ellipticals$^{\bf\rm 39}$.  In a
cluster, one of these ``cannablises " its neighbors to become the
giant central ellpitical$^{\bf\rm 40}$.  The dwarf ellipticals are
created by the harassment of low luminosity spirals.  Harassment has
the potential to change any internal property of a galaxy within a
cluster including the gas distribution and content, the orbital
distribution of stars and the overall shape.  Our first examination
has only touched on some of the most dramatic changes, the
phenomenology of harassment promises to be even richer than that of
merging and cannibalism.

\oneskip

\noindent{\bf Acknowledgments} \ This research was funded by NASA
through the LTSA, ATP and HPCC/ESS programs.
\oneskip

\vfil\eject
\noindent{\bf References}

\pp 1. Dressler A, Oemler A., Gunn J.E. \& Butcher H.  1993,
{\it Astrophys.J. Lett.}, {\bf 404}, L4-6.

\pp 2. \ Dressler A. 1980, {\it Astrophys.J.}, {\bf 236}, 351-65.

\pp 3. \ Butcher H. \& Oemler A. 1978, {\it Astrophys.J.}, {\bf 219}, 18-33.

\pp 4. \ Butcher H. \& Oemler A. 1984, {\it Astrophys.J.}, {\bf 285}, 426-38.

\pp 5. \ Dressler A, Oemler A., Butcher H. \& Gunn J.E.  1994a,
{\it Astrophys.J.}, {\bf 430}, 107-20.

\pp 6. \ Dressler A, Oemler A., Sparks W.B. \& Lucas R.A. 1994b,
{\it Astrophys.J.Lett.}, {\bf 435}, L23-6.

\pp 7. \ Couch W.J., Ellis R.S., Sharples R. \& Smail I. 1994, {\it
Astrophys.J.},
{\bf 430}, 121-38.

\pp 8. \ Barger, A. J., Aragon-Salamanca, A., Ellis, R. S., Couch, W. J.,
Smail,
I. and Sharples, R. M. 1995, {\it Mon.Not.R.Astr.Soc.}, in press.

\pp 9. \ Griffiths R.E. \etal 1994, {\it Astrophys.J.Lett.}, {\bf 435}, L19-22.

\pp 10. Schechter P. 1976, {\it Astrophys.J.}, {\bf 203}, 297-306.

\pp 11. Binggeli B., Tammann G.A. \& Sandage A. 1987, {\it Astron.J.}, {\bf
94}, 251-77.

\pp 12. Thomspon L.A.\& Gregory S.A. 1993, {\it Astron.J.}, {\bf 106},
2197-212.

\pp 13. Kauffmann G. 1995, {\it Mon.Not.R.astr.Soc.}, {\bf 274}, 153-60.

\pp 14. Icke V. 1985, {\it Astron.Astrophys.}, {\bf 144}, 115-23.

\pp 15. Miller R.H. 1988, {\it Comment. Astrophys.}, {\bf 13}, 1-11.

\pp 16. Dressler A. \& Gunn J.E. 1983, {\it Astrophys.J.}, {\bf 270}, 7-19.

\pp 17. Evrard A.E. 1991, {\it Mon.Not.R.astr.Soc.}, {\bf 248}, 8p-10.

\pp 18. Byrd, G. and Valtonen-M. 1990, {\it Astrophys.J.}, {\bf 350}, 89-94.

\pp 19. Valluri M. 1993, {\it Astrophys.J.}, {\bf 408}, 57-70.

\pp 20. Oemler, A., Dressler, A. and Butcher, H. R. 1995,
{\it Astrophys.J.}, submitted.

\pp 21. Hernquist L. \& Katz N. 1989, {\it Astrophys.J.Supp.}, {\bf 70},
419-46.

\pp 22. Tully R.B. \& Fisher J.R. 1977, {\it Astro.Ap.}, {\bf 54}, 661.

\pp 23. Faber S.M. \& Jackson R.E. 1976, {\it Astrophys.J.}, {\bf 204}, 668-83.

\pp 24. Moore B., Katz N. \& Lake G. 1995, {\it Astrophys.J.}, in press.

\pp 25. White S.D.M. \& Rees M.J. 1978, {\it Mon.Not.R.Astr.Soc.}, {bf 183},
341-58.

\pp 26. van der Marel, R. P. 1991, {\it Mon.Not.R.Astr.Soc.}, {\bf 253},
710-26.

\pp 27. Sandage, A. and Bedke, J. 1994, {\it The Carnegie Atlas of
Galaxies}, Carnegie Institute of Washington.

\pp 28. Combes F., Dupraz C., Casoli F. \& Pagani L. 1988, {\it
Astron.Astrophys.},
{\bf 203}, L9-12.

\pp 29. Kormendy, J. and Bender, R. 1995, in {\it ESO/OHP Symposium
on Dwarf Galaxies,}, ed. G. Meylan and P. Prugniel, European
Southern Observatory, in press.

\pp 30. Ferguson H.C. \& Binggeli B. 1994, {\it Astronomy and Astrophysics
Review},
{\bf 6}, 67-122.

\pp 31. Barnes J.E. \& Hernquist L. 1992, {\it Nature}, {\bf 360}, 715-17.

\pp 32. Mirabel I.F., Dottori H. \& Lutz D. 1992, {\it Astron.Astrophys.}, {\bf
256}, L19-22.

\pp 33. Gudehus D.H. 1989, {\it Astrophys.J.}, {\bf 340}, 661-5.

\pp 34. Tyson J.A. \& Fischer P. 1995, {\it Astrophys.J.Lett.}, in press.

\pp 35. Bahcall, J.N., Kirkhakos S. \& Schneider D.P. 1994, {\it Astrophys.J.
Lett.},
{\bf 435}, L11-14.

\pp 36. Disney M.J. \etal 1995, {\it Nature}, {\bf 376}, 150-3.

\pp 37. French H.B. \& Gunn J.E. 1983, {\it Astrophys.J.}, {\bf 269}, 29-34.

\pp 38. Lake G., Katz N. \& Moore B. 1995, {\it Astrophys.J.}, in press.

\pp 39. Toomre A. 1977, In {\it The Evolution of Galaxies and Stellar
Populations},
ed. B.M. Tinsley \& R.B. Larson, p. 401-17. Yale University Observatory.

\pp 40. Ostriker J.P. \& Hausman M.A. 1977, {\it Astrophys.J.Lett.}, {\bf 217},
L125-9.

\end{document}